\documentclass[11pt]{article}
\usepackage{epsf}

\usepackage{epsfig,graphics}
\usepackage {graphicx}
\usepackage{graphbox}
\usepackage {epsfig}
\usepackage {subfigure}
\usepackage {tabularx} 
\usepackage{rotate}	
\usepackage{slashed}

\usepackage{amsmath}
\usepackage{amsfonts}
\usepackage{amssymb}
\usepackage{cite}
\usepackage{color}
\usepackage{bbm}

\usepackage[utf8]{inputenc} 
\usepackage{fancyhdr}
\usepackage{hyperref}

\newcolumntype{M}[1]{>{\centering\arraybackslash}m{#1}}
\newcolumntype{N}{@{}m{0pt}@{}}

\newcommand{\bmat}{\left(\begin{array}}
\newcommand{\emat}{\end{array}\right)}

\def\preal{{\rm Re\,}}
\def\pim{{\rm Im\,}}

\def\yzero{\smash{\hbox{$y\kern-4pt\raise1pt\hbox{${}^\circ$}$}}}

\def\beq{\begin{equation}}
\def\eeq{\end{equation}}
\def\beqa{\begin{eqnarray}}
\def\eeqa{\end{eqnarray}}

\def\-{\hphantom{-}}

\def\s2{\frac{1}{\sqrt2}}

\def\beq{\begin{equation}}
\def\eeq{\end{equation}}
\def\beqa{\begin{eqnarray}}
\def\eeqa{\end{eqnarray}}

\def\IF{\relax{\rm I\kern-.18em F}}
\def\II{\relax{\rm I\kern-.18em I}}
\def\IP{\relax{\rm I\kern-.18em P}}
\def\IC{\relax\hbox{\kern.25em$\inbar\kern-.3em{\rm C}$}}
\def\IR{\relax{\rm I\kern-.18em R}}

\def\Dsl{\,\raise.15ex\hbox{/}\mkern-13.5mu D} 

\def\bulk{\mathrm{bulk}}
\def\DBI{\mathrm{DBI}}
\def\CS{\mathrm{CS}}

\catcode`\@=11   
\newdimen\@rotdimen
\newbox\@rotbox  

\def\@vspec#1{\special{ps:#1}}
\def\@rotstart#1{\@vspec{gsave currentpoint currentpoint translate
   #1 neg exch neg exch translate}}
\def\@rotfinish{\@vspec{currentpoint grestore moveto}}
\def\@rotr#1{\@rotdimen=\ht#1\advance\@rotdimen by\dp#1%
   \hbox to\@rotdimen{\hskip\ht#1\vbox to\wd#1{\@rotstart{90 rotate}%
   \box#1\vss}\hss}\@rotfinish}
\def\@rotl#1{\@rotdimen=\ht#1\advance\@rotdimen by\dp#1%
   \hbox to\@rotdimen{\vbox to\wd#1{\vskip\wd#1\@rotstart{270 rotate}%
   \box#1\vss}\hss}\@rotfinish}%
\def\@rotu#1{\@rotdimen=\ht#1\advance\@rotdimen by\dp#1%
   \hbox to\wd#1{\hskip\wd#1\vbox to\@rotdimen{\vskip\@rotdimen
   \@rotstart{-1 dup scale}\box#1\vss}\hss}\@rotfinish}%
\def\@rotf#1{\hbox to\wd#1{\hskip\wd#1\@rotstart{-1 1 scale}%
   \box#1\hss}\@rotfinish}%
\def\rotate{\@ifnextchar[{\@rotate}{\@rotate[l]}}
\def\@rotate[#1]#2{\setbox\@rotbox=\hbox{#2}\@nameuse{@rot#1}\@rotbox}

\catcode`\@=12

\topmargin
-1.5cm
\textwidth
15.5cm
\textheight
23.5cm
\oddsidemargin
0.7cm
\evensidemargin
0.7cm

\begin{document}

\makeatletter
\@addtoreset{equation}{section}
\makeatother
\renewcommand{\theequation}{\thesection.\arabic{equation}}
\pagestyle{empty}
\vspace{-0.2cm}
\rightline{ IFT-UAM/CSIC-20-69}
\vspace{1.2cm}
\begin{center}


\LARGE{A Note on Membrane Interactions and the Scalar potential\\ [13mm]}
  \large{Alvaro Herr\'aez
   \\[6mm]}
\small{

 {\em Departamento de F\'{\i}sica Te\'orica
and Instituto de F\'{\i}sica Te\'orica UAM/CSIC,\\[-0.3em]
Universidad Aut\'onoma de Madrid,
Cantoblanco, 28049 Madrid, Spain} \\[5mm]
}
\small{\bf Abstract} \\[6mm]
\end{center}
\begin{center}
\begin{minipage}[h]{15.22cm}
We compute the tree-level potential between two parallel $p$-branes due to the exchange of scalars, gravitons and $(p+1)$-forms. In the case of BPS membranes in 4d $\mathcal{N}=1$ supergravity, this provides an interesting reinterpretation of the classical Cremmer et al. formula for the F-term scalar potential in terms of scalar, graviton and 3-form exchange.  In this way, we present a correspondence between the scalar potential at every point in scalar field space and a system of two interacting BPS membranes. This could potentially lead to interesting implications for the Swampland Program by providing a concrete way to relate conjectures about the form of scalar potentials with conjectures regarding the spectrum of charged objects.
\end{minipage}
\end{center}
\newpage
\setcounter{page}{1}
\pagestyle{plain}
\renewcommand{\thefootnote}{\arabic{footnote}}
\setcounter{footnote}{0}

\tableofcontents

\section{Introduction}

It is well known that the 4d cosmological constant can be interpreted in terms of field strengths of  3-forms. Even though they do not propagate additional degrees of freedom,  they can acquire non-vanishing vevs and give rise to a  cosmological constant contribution. For this reason, 3-forms have been used in trying to solve the cosmological constant problem as in  \cite{pioneros1,pioneros2,pioneros3,pioneros4,pioneros5}. More specifically, considering the membranes to which a 3-form naturally couples provides a mechanism for the cosmological constant to change when a membrane is crossed, as considered originally in \cite{BT, BT2}, and also in \cite{BP} within the context of String Theory. In fact, this relation works and has been studied not only for constant contributions but for more general scalar potentials including axions in \cite{morerecent2,morerecent3,morerecent4,morerecent5, KS, KLS}. In String Theory, this has also been explored \cite{Dudas:2014pva,Escobar:2015ckf,imuv,Carta:2016ynn,Garcia_Valdecasas:2016voz,Valenzuela:2016yny,Blumenhagen:2017cxt} and in the context of type II compactifications with fluxes, it was shown in \cite{Bielleman:2015ina, Herraez:2018vae}  that the complete F-term flux potential can be expressed, after integrating out the 4-forms, as
\begin{equation}
V=\dfrac{1}{2}Z^{AB}Q_AQ_B,
\end{equation}
where $Z^{AB}$ includes the field dependence and can be obtained from the kinetic terms of the 3-forms. The $Q_A$ give the couplings of the corresponding 4-form. The fact that this $\mathcal{N}=1$ potential can be expressed completely in terms of 3-forms, which naturally couple to membranes, suggests that a direct relation between these objects can be drawn. In particular, the potential being a bilinear in the charges reminds of a membrane-membrane interaction, and this is precisely the relation that we study in this note. We particularize for the case two interacting BPS membranes in 4d and point out a correspondence between their different interactions and the different terms in the Cremmer et al. $\mathcal{N}=1$ F-term scalar potential (see eq. \eqref{Cremmeretal}). Moreover, no obstruction to the application of the same logic to codimension 1 branes in higher dimensions is expected. 

This correspondence is interesting by itself, but it can also be useful in the context of the Swampland \cite{swampland} (see also \cite{vafafederico, review} for interesting reviews), since it could provide a precise setup in which two types of conjectures may be related. On the one hand,  hypothesis about the generic properties of the scalar potentials that can arise in  QG have been the subject of a considerable study recently, including the idea that metastable de Sitter space cannot exist  \cite{dS1, Krishnan, dS3} or has to be sufficiently short-lived \cite{TCC}, or the suggestion that stable non-susy AdS \cite{Ooguri:2016pdq}, as well as scale separated AdS, belong to the Swampland \cite{lpv, Gautason:2015tig}. On the other hand, a lot of progress has been made in clarifying conjectures about properties of the spectrum of QG theories, like the Weak Gravity Conjecture \cite{WGC} or the Swampland Distance Conjecture \cite{distance}. Besides, several connections between apparently different Swampland Conjectures have been gradually uncovered (see e.g. \cite{review, timo2, Gendler:2020dfp,Andriot:2020lea} and references therein), realizing the idea of a network of interconnected conjectures, instead of a set of unrelated statements. In this context, the precise correspondence between the scalar potential and the interactions of two membranes could provide new ways to relate the restrictions on the scalar potentials with the properties of the membranes.\footnote{Some connections between these two kinds of statements have already been pointed out in \cite{PaltiTalkSP19}, and also along the lines of potentials arising from integrating out towers of states in \cite{dS3, review}.}

In section \ref{section2}, we review the classical field theory calculation of the interaction between $p$-branes due to scalar, graviton and $(p+1)$-forms exchange, and use it to interpret a particular version of the WGC. Section \ref{section3} presents the main result of this note, namely the correspondence between the different pieces of the Cremmer et al. scalar potential and the different interactions between a pair of flat BPS membranes. We leave the summary and outlook for the final section.

\emph{Note added:} When finalizing this note we were informed that a related paper is about to appear \cite{LMMV}, also pointing out the interpretation of the $\mathcal{N}=1$ F-term potential as a no-force condition for the membranes and studying implications for the swampland conjectures.

\section{Interactions between D$p$-branes in D Dimensions}
\label{section2}
We begin by reviewing the field theory calculation of the tree-level potential between two parallel, infinite $p$-branes in $D$-dimensional Minkowski space \cite{Polchinski:1998rq,Polchinski:1998rr}.\footnote{In stringy terms, this corresponds to closed string exchange, which in the field theory limit reduces to the exchange of scalars and gravitons from the NSNS sector and $(p+1)$-forms from the RR sector \cite{Polchinski:1995mt}.} The goal of this section is to review this calculation in detail to fix the conventions and explicitly keep track of the units, which is crucial for comparison with Swampland Conjectures. For concreteness, let us consider a $D$-dimensional generalization of the low energy effective action of type II supergravity and include source terms corresponding to $Dp$-branes with charge $Q_p$  and tension $\tilde{T}_p$ (in string units) \cite{Polchinski:1998rq,Polchinski:1998rr, BOOK}.

\begin{equation}
\label{Stotal}
    S=S_{\bulk}+S_{\DBI}+S_{\CS},
\end{equation}
where the different pieces take the form:
\begin{align}
\label{Sbulk}
    S_{\bulk}  = &   \displaystyle \dfrac{1}{2 \tilde{\kappa}_{D}^{2}} \, \int d^{D} x \sqrt{-\tilde{g}} \, e^{-2 \tilde{\phi}} \, \left(\tilde{R}+4(\partial \tilde{\phi})^{2}\right)-\frac{1}{4 \tilde{\kappa}_{D}^{2}} \int  G_{p+2} \wedge \tilde{\star} G_{p+2} \, , \\
    \label{SDBI}
      S_{\DBI} = & -\tilde{T}_p    \int_{WV} d^{p+1}\xi \, e^{-\tilde{\phi}} \,  \left[-\det \left( \tilde{g}_{ab}+\mathcal{F}_{ab} \right) \right]^{1/2}\, , \\ 
      \label{SCS}
         S_{\CS} = &  Q \int_{WV} C_{p+1} \, .
\end{align}

This action is expressed in the string frame and the quantities are measured in string units.\footnote{We denote all quantities in the string frame by tildes and use latin letters to refer to worldvolume indices whereas greek letters denote spacetime indices.} Separating the dilaton into a background and a dynamical part, that is, $\tilde{\phi}=\bar{\phi}+\phi$ we can define the quantities $\kappa^2_D=\tilde{\kappa}^2_D e^{2\bar{\phi}}$ and $T_p=\tilde{T}_p e^{-\bar{\phi}}$, which are the gravity coupling constant and the effective tension of the membrane, respectively. Then, upon performing a Weyl rescaling of the metric $\tilde{g}_{\mu\nu}=e^{\frac{4}{D-2}\phi}g_{\mu \nu}$ we obtain the following expressions for $S_{\bulk}$ and $S_{\DBI}$ in the Einstein frame ($S_{\CS}$ does not change since it is a topological term, independent of the bulk metric).
\begin{align}
\label{SbulkE}
         \displaystyle S_{\bulk} = &   \dfrac{1}{2 \kappa_{D}^{2}} \, \int d^{D} x \sqrt{-g} \, \left(R+\dfrac{4}{D-2}(\partial \phi)^{2}\right)-\frac{e^{2\bar{\phi}}}{4 \kappa_{D}^{2}} \int  e^{-(\frac{4p-2D+8}{D-2})\phi}\, G_{p+2} \wedge \star G_{p+2}\,  , \\ 
       \label{SDBIE}
      S_{\DBI} = & -T_p   \displaystyle \int_{WV} d^{p+1}\xi \, \exp{\left[\left(\frac{2p-D+4}{D-2} \right) \phi \right]} \,  \sqrt{-\det \left( g_{ab} \right)}\, .
\end{align}
We use $\kappa_D$ in this section, since it is more suitable for perturbative calculations but keep in mind that the Planck mass is given by $\kappa_D^{-2}=M_p^{D-2}$. From here, we can compute the tree-level scalar, graviton and $(p+1)$-form exchange between two membranes. The propagators associated to the graviton, scalar and $(p+1)$-form can be obtained from $S_{\bulk}$ and the interaction vertices between the brane and the fields from $S_{\DBI}$ (for the scalar and graviton) and $S_{\CS}$ (for the $(p+1)$-form). 

\subsection{The Scalar plus Graviton Interaction}

\begin{figure}[tb]

    \begin{center}

        \includegraphics[align=c, height=58pt]{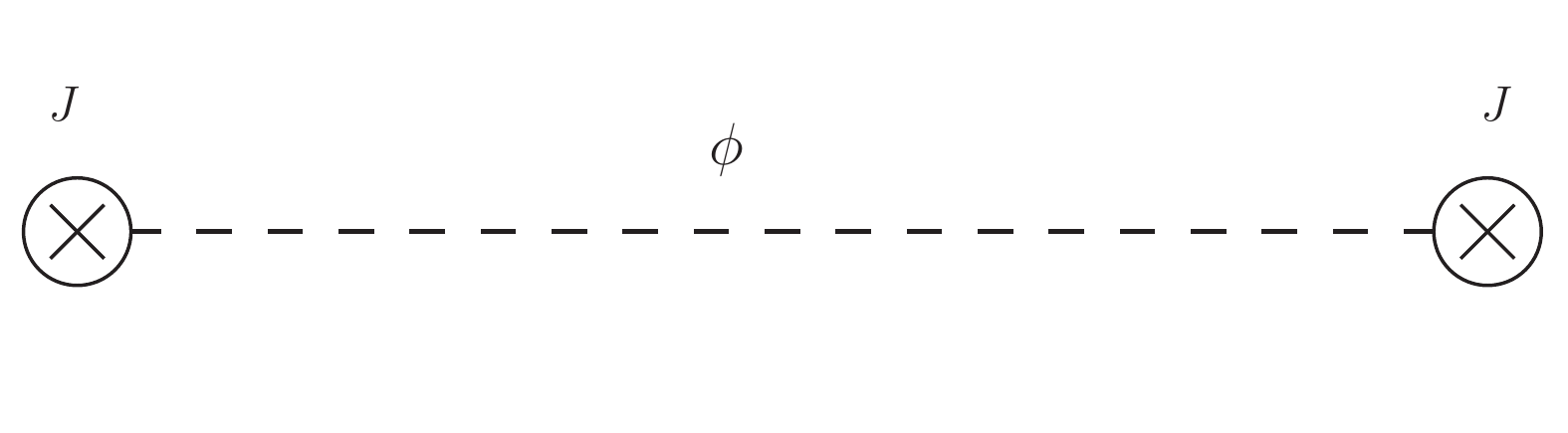}
        \label{fig:scalarexchange}
       $+$    
        \includegraphics[align=c, height=58pt]{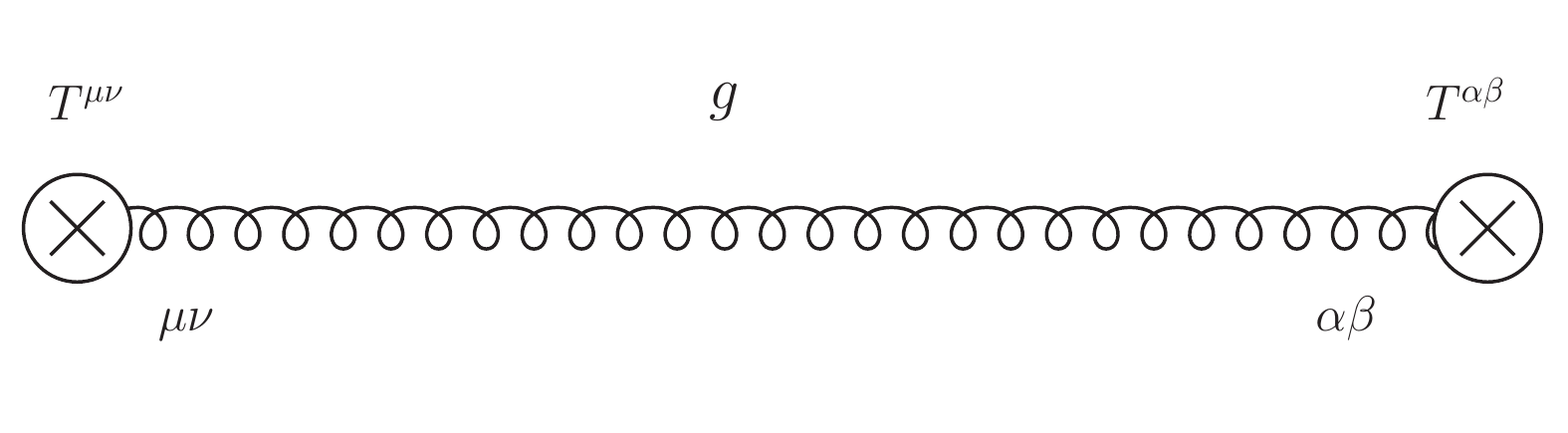}
        \label{fig:gravitonexchange}
        \caption{Diagramatic representation of the tree-level scalar and  graviton exchange between two $p$-branes}
       \label{fig:NSexchange}
     \end{center}
\end{figure}

Let us begin with the interaction due to the scalar and graviton exchange, which corresponds to the diagrams in fig. \ref{fig:NSexchange}

Expanding the metric as a background plus a perturbation as $g_{\mu \nu}=\bar{g}_{\mu \nu}+ \kappa_D h_{\mu \nu}$ we compute
\begin{equation}
    \sqrt{-\det(g_{ab})}=\sqrt{|\bar{g}_{ab}|}\left(1+\dfrac{\kappa_D}{2}\bar{g}^{ab}h_{ab}+... \right),
    \label{detexpansion}
\end{equation}
where we have defined $|\bar{g}_{ab}|=-\det(\bar{g}_{ab})$. Using this expression and working in De Donder gauge \cite{Alvarez:2020zul}, we can expand the Ricci scalar to obtain the graviton propagator in momentum space \cite{Polchinski:1998rq,Polchinski:1998rr}:
\begin{equation}
    \langle h_{\mu \nu} h_{\alpha \beta} \rangle  =-\dfrac{2 i}{k^2}\left( \bar{g}_{\mu \alpha} \bar{g}_{\nu \beta}+\bar{g}_{\mu \beta} \bar{g}_{\nu \alpha}-\dfrac{2}{D-2}\bar{g}_{\mu \nu} \bar{g}_{\alpha \beta} \right) .
\end{equation}
Additionally, the scalar propagator takes the form
\begin{equation}
     \langle \phi \phi \rangle = -\dfrac{i \kappa_D^2}{k^2}\dfrac{(D-2)}{4}.
\end{equation}
 
We calculate the relevant vertices by expanding $S_{\DBI}$  to obtain
\begin{equation}
\begin{split}
   S_{\DBI} = & -T_p    \int_{WV} d^{p+1} \xi\sqrt{|\bar{g}_{ab}|}  -T_p    \int_{WV} d^{p+1} \xi\sqrt{|\bar{g}_{ab}|} \left( \dfrac{2p-D+4}{D-2}\right) \phi +  \\ & - T_p    \int_{WV} d^{p+1} \xi\sqrt{|\bar{g}_{ab}|} \, \dfrac{\kappa_D}{2}\bar{g}^{ab}h_{ab}+ \ldots
    \end{split}
\end{equation}
where the first term gives the usual contribution from the embedding of the worldvolume in the $D$-dimensional spacetime, the second gives the interaction between one scalar and the source and the third gives the interaction between the graviton and the energy-momentum tensor of the brane. The ellipsis indicate interactions of the source with more than one field, which are not relevant for our computation. For a flat membrane in Minkowski space we can choose a set of coordinates such that $x^0=\xi^0, x^1=\xi^1 \,  ... \ x^p=\xi^p $ and $x^{p+1}=...=x^{D-1}= \mathrm{const}$, impliying 
 $g_{ab}=g_{\mu \nu} \delta_a^\mu \delta_b^\nu$.

The worldvolume metric then takes the form of the corresponding $(p+1)\times (p+1)$ block of the background $D$-dimensional metric. Using this we obtain the following Feynman rules for the relevant vertices
\begin{equation}
\includegraphics[align=c, height=48pt]{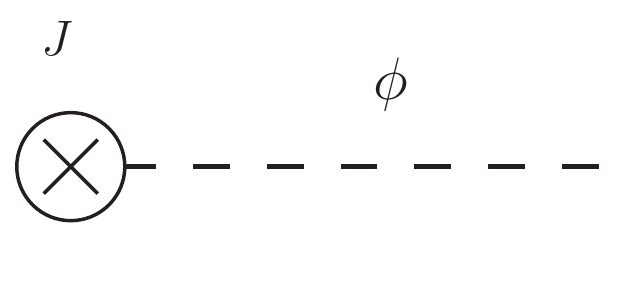}
        \label{fig:scalarvertex} \  = \  i T_p \sqrt{|\bar{g}_{ab}|} \left( \dfrac{2p-D+4}{D-2} \right),
\end{equation}
\begin{equation}
\includegraphics[align=c, height=48pt]{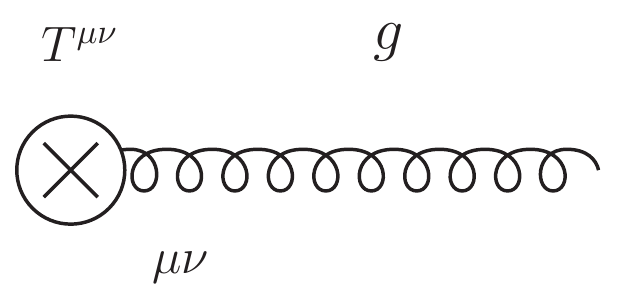}
        \label{fig:gravitonvertex} \  = \  i T_p \sqrt{|\bar{g}_{ab}|} \dfrac{\kappa_D}{2}\bar{g}^{ab} \delta_a^\mu \delta_b^\nu. \qquad 
\end{equation}
 
 The amplitude for the scalar and graviton interaction given in fig. \ref{fig:NSexchange} then yields
 \begin{equation}
 \label{As+ggen}
     \mathcal{A}_{s+g}=\dfrac{T_p^2 |\bar{g}_{ab}| \kappa_D^2}{k^2} \left\{ \dfrac{(2p-D+4)^2}{4(D-2)}+\dfrac{(D-p-3)(p+1)}{D-2}  \right\}= \dfrac{T_p^2 |\bar{g}_{ab}| \kappa_D^2}{k^2} \left\{ \dfrac{D-2}{4}\right\},
 \end{equation}
where $k$ indicates the momentum in the directions perpendicular to the membranes. The first term corresponds to the scalar interaction and the second to the graviton exchange. The final result is independent of $p$, as can be seen in the last step, but we will keep both terms in order to keep track of the scalar and the graviton pieces separately. Since we will be mainly interested in codimension 1 objects (i.e. $p=D-2$), let us particularize for that case, in which the amplitude takes the form:
\begin{equation}
     \mathcal{A}_{s+g}=\dfrac{T_p^2 |\bar{g}_{ab}| \kappa_D^2}{k^2}\left\{ \dfrac{D^2}{4(D-2)}-\dfrac{D-1}{D-2}  \right\}.
     \label{As+g}
\end{equation}

In this case, the scalar contribution is always positive (i.e. attractive) but the contribution from the graviton exchange becomes negative, yielding a repulsive force, which only occurs in this particular case of codimension 1 objects. The potential between the two membranes can be calculated by taking the Fourier transform of this amplitude, and for the codimension 1 objects, it grows linearly with the distance, implying a force that does not depend on the distance.

 \subsection{The q-form Interaction}
 
 \begin{figure}[tb]
  \label{fig:qformexchange}
   \begin{center}
 \includegraphics[align=c, height=48pt]{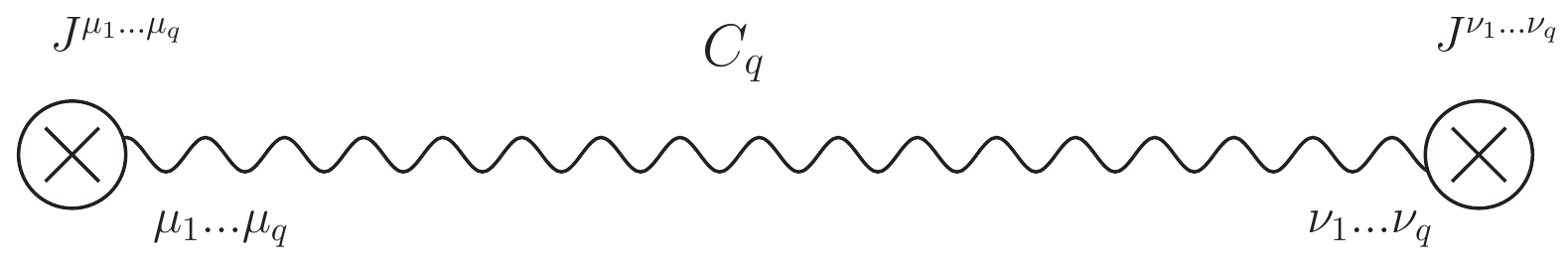}
      
          \caption{Diagramatic representation of the tree-level $q$-form exchange between two $(q-1)$-branes}  
     \end{center}
\end{figure}

 The $q$-form interaction between two $(q-1)$-branes corresponds to the diagram in fig. \ref{fig:qformexchange}. To compute it we need the propagator of the $q$-form and its coupling to the membrane. We begin with the propagator and, in order to obtain the tensor structure, we first consider a canonically normalized $q$-form kinetic term and after we include the overall (background-field dependent) prefactors appearing in \eqref{SbulkE}. We consider then the kinetic part of the action of a q-form $A_q=\dfrac{1}{q!}A_{\mu_1 ...\mu_q }dx^{\mu_1}\wedge ... \wedge dx^{\mu_q}$, which takes the form
 \begin{equation}
     S_{\mathrm{q,\, kin}}=-\dfrac{1}{2 (q+1)!}\int d^D x \sqrt{-g} F_{\mu_1...\mu_{q+1}}F^{\mu_1...\mu_{q+1}},
 \end{equation} 
 with $F_{\mu_1...\mu_{q+1}}=(q+1) \partial_{[\mu_1}A_{\mu_2 ...\mu_{q+1}] } $. \footnote{Square brackets indicate antisymmetrization with a normalization factor of $1/q!$ in front, so that for an antisymmetric $q$-form we have $A_{\mu_1 ...\mu_q }=A_{[\mu_1 ...\mu_q ]}\, $.} Integrating by parts and massaging the Lagrangian we obtain
 \begin{equation}
   \label{qkinnogf}
     \begin{split}
           \mathcal{L}_{\mathrm{q,\, kin}} = & \dfrac{1}{2 q!}A_{[\mu_1...\mu_q]}\left\{  -  \delta_{\nu_1}^{[\mu_1}...\delta_{\nu_q}^{\mu_q]} \partial_\rho \partial^\rho + q\, \delta_{\rho}^{\mu_1}\delta_{[\nu_1}^{\sigma} \delta^{\mu_2}_{\nu_2}...\delta_{\nu_q]}^{\mu_q} \partial_\sigma \partial^\rho \right\}A^{[\nu_1...\nu_q]}.
     \end{split}
 \end{equation}
In order to obtain the propagator, we need to invert the second variation of this part of the action with respect to the $q$-form field. To do so we choose a generalization of the Lorentz gauge \cite{Luscher} for $q$-forms, namely $\partial_\mu A^{[\mu \nu_2...\nu_q]}=0$,
and implement it by means of the following gauge fixing term in the Lagrangian  $\mathcal{L}_{\mathrm{gf}}\sim \partial_\rho A^{[\rho  \nu_2...\nu_q]}\partial^\sigma A_{[\sigma \nu_2...\nu_q]}$. By adjusting the constant in front we can cancel the second term in the last line of eq. \eqref{qkinnogf}, so that we are left with the task of inverting the operator
\begin{equation}
    K^{\mu_1...\mu_n}_{\nu_1...\nu_n}=\dfrac{1}{q!}\left(-  \delta_{[\nu_1}^{[\mu_1}...\delta_{\nu_q] }^{\mu_q]} \partial_\rho \partial^\rho \right),
\end{equation}
which yields the propagator (in momentum space)
\begin{equation}
    \langle A^{\nu_1...\nu_n} A_{\mu_1...\mu_n}  \rangle = i \, \dfrac{q!}{k^2}   \delta_{[\nu_1}^{[\mu_1}...\delta_{\nu_q] }^{\mu_q]}  .
\end{equation}
Note that this is normalized in such a way that the propagator from any independent component to itself (or any antisymmetric permutation thereof) coincides with the propagator of a scalar degree of freedom, as expected for a field with a canonical kinetic term. In order to calculate the propagator of the $C_{p+1}$ from eq. \eqref{Sbulk} we just need to rescale the calculated propagator to account for the overall factors that prevent the kinetic term from being canonically normalized, obtaining\footnote{Notice that only the$\bar{\phi}$ dependent part of the exponential factor eq. \eqref{SbulkE} contributes to the 2-point function, as the other exponential piece disappears at leading order in $\phi$.}
\begin{equation}
\label{qformprop}
 \langle C^{\nu_1...\nu_q} C_{\mu_1...\mu_q}  \rangle =2i \, \dfrac{ \kappa_D^2\,  e^{-2\bar{\phi}}}{k^2} q! \delta_{[\nu_1}^{[\mu_1}...\delta_{\nu_q] }^{\mu_q]}  
\end{equation}

To obtain the $q$-form brane vertex, we use eq. \eqref{SCS}, which in components reads
\begin{equation}
    S_{CS}=\dfrac{Q}{q!}\int_{WV} d^q \xi\, C_{\mu_1...\mu_q} \left( \dfrac{\partial x^{\mu_1}}{\partial \xi^{a_1}}...\dfrac{\partial x^{\mu_q}}{\partial \xi^{a_q}} \right) \epsilon^{a_1...a_q}.
\end{equation}
For a flat brane we obtain the following Feynman rule for the vertex 
\begin{equation}
\includegraphics[align=c, height=48pt]{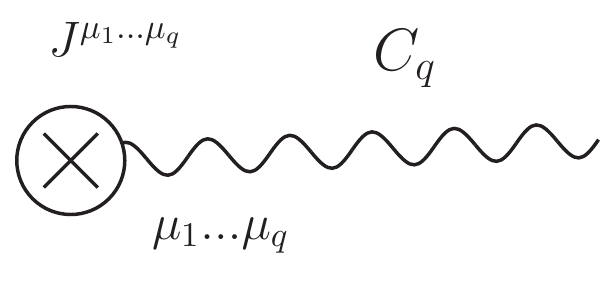}
        \label{fig:qformvertex} \  = \  i \dfrac{Q}{q!}\epsilon^{a_1...a_q} \delta_{a_1}^{[\mu_1}...\delta_{a_q }^{\mu_q]}.
\end{equation}
The $q$-form exchange of fig. \ref{fig:qformexchange} yields the following amplitude
\begin{equation}
    \mathcal{A}_C=  -\dfrac{Q^2}{(q!)^2}\dfrac{2\, \kappa_D^2 \,  e^{-2\bar{\phi}}}{k^2}\, q!\, \epsilon^{a_1...a_q} \bar{g}_{a_1 b_1}...\bar{g}_{a_q b_q}\epsilon^{b_1...b_q}= -\dfrac{2 Q^2 |\bar{g}_{ab}| \kappa_D^2 e^{-2\bar{\phi}}}{k^2},
    \label{AC}
\end{equation}
where we have used $\epsilon^{a_1...a_q} \bar{g}_{a_1 b_1}...\bar{g}_{a_q b_q}\epsilon^{b_1...b_q}= q!\, |\bar{g}_{ab}|$ in the last step. Notice that this depends on the rank of the form and the spacetime dimension only implicitly via $g_{ab}$ and $k^2$, but not explicitly as in the scalar and graviton exchange. 

Then, the three contributions cancel, and no net force is felt by the branes if
\begin{equation}
\label{equality}
\left\{ \dfrac{D^2}{4(D-2)}-\dfrac{D-1}{D-2}  \right\}T_p^2 =2 Q^2 e^{-2\bar{\phi}}.
\end{equation}
This is what happens for single $Dp$-branes in 10 dimensions, which are BPS objects, upon substitution of the corresponding tensions and charges (i.e.  $Q=\tilde{T}_p=T_p e^{\bar{\phi}}$). This cancellation is precisely expected for BPS objects, since they feel no net force. For the case of codimension 1 branes, the attractive contribution from the scalars compensates the repulsive force from graviton and $q$-form exchange. 

\subsection{Relation with the WGC for $q$-forms in the presence of dilaton-like couplings}

In this section, we make a small detour from the main goal of this note and explore the consistency of this calculation with the WGC. The aforementioned case of single D$p$-branes in ten dimensions represents a particular realization of the general claim that WGC inequality is saturated occurs for BPS objects in supersymmetric theories\cite{Ooguri:2016pdq}. More generally, we can recover the results for the extremal form of the WGC for $q$-forms in the presence of dilaton-like couplings proposed in \cite{WGC16} by requiring that eqs. \eqref{As+g} and \eqref{AC} cancel each other out.\footnote{This is just a manifestation of the special case in which arguments about force cancellation and extremality of black hole solutions coincide (for a more detailed discussion on the (in)equivalence of these two approaches see \cite{Palti, timo2, Heidenreich:2019zkl, Gendler:2020dfp}).} This form of the WGC can be expressed, using our notation, as the following inequality
\begin{equation}
    \kappa_D^2 \left[\frac{\alpha^{2}}{2}+\frac{q(D-q-2)}{D-2}\right] T^{2} \, \leq \, e^2 \,  Q^{2} \, 
    \label{WGCqforms}
\end{equation}
This expression is obtained from the extremality bound of black branes in theories with gravity, a $q$-form and a dilaton-like scalar \cite{Horowitz:1991cd}. The r.h.s. comes from the $q$-form interaction, and in our case, the gauge coupling can be read from eq. \eqref{SbulkE} and it equals  $e^2=2 \kappa^2_D e^{-2\bar{\phi}}$. This matches exactly the r.h.s. of \eqref{equality}. The second term of the l.h.s. comes from the gravitational interaction, which depends on the spacetime dimensions and the rank of the form, and it can be checked that it matches the second term in the l.h.s. of \eqref{equality} upon substitution of $q=p-1$. Finally, the first term on the l.h.s. corresponds to the scalar interaction, whose coupling constant can be extracted from the kinetic terms of the $q$-form. In particular, for a theory with a $q$-form and a \textit{conventionally normalized} scalar field $\phi$, a dilaton-like coupling  is characterized by a term of the following form in the Lagrangian $\mathcal{L}_{\mathrm{q,\, kin}} \sim e^{-\vec{\alpha}\cdot \vec{\phi}} F^{2}$ (see  \cite{WGC16}), where $\alpha$ is the dilaton coupling constant. In our model, we need to identify the $\alpha$ from the kinetic terms for the $q$-forms in the Einstein frame, given by eq. \eqref{SbulkE}. From the exponential, we can read the coupling to the dilaton $\phi$, and after conventionally normalizing its kinetic term via $\phi_c=\sqrt{\frac{8}{D-2}}\phi$ we obtain a kinetic term of the form described above with 
\begin{equation}
\label{alpha}
    \dfrac{\alpha^2}{2}=\dfrac{(2p-D+4)^2}{4D-8}.
\end{equation}
This exactly resembles the contribution from the scalar exchange in eq. \eqref{equality}. As advertised, this extremal form of the WGC, which was originally formulated from extremality arguments for BH's, can be obtained from the requirement that the force between charged objects cancels exactly (see also \cite{Palti, timo2, Heidenreich:2019zkl, Gendler:2020dfp} for more on this approach or \cite{Gonzalo:2020kke} for an alternative approach to the WGC in terms of pair production).

Let us remark that the action considered at the beginning of this section is really meaningful in $D=10$, since in this case it is a piece of the 10d type II effective action coupled to D$p$-branes. The extension to general $D$ serves for illustrative purposes, since it allows to capture some aspects of the typical $D$-dimensional effective actions, like the Lorentz index structure of the propagators. However, eq. \eqref{alpha} should not be taken as a general expression, but rather as a check for the case in which a $D$-dimensional action can be recast into the form given in eqs. \eqref{Stotal}-\eqref{SDBIE}. In particular, if we focus on  particles in 4d (i.e. $p=0$ and $D=4$), this action could capture the coupling of particles coming from D$p$-branes wraping internal p-cycles to the saxionic fields in the complex structure sector of type IIA compactifications, which is known to vanish \cite{irene, Font:2019cxq}. In order to capture other couplings like the ones that would come from the K\"ahler sector in type IIA compactifications, or axions, one should take a more general 4d effective action, but this is not the main goal of this section so we will not elaborate more on this. Let us just mention that the kind of couplings that can be encoded in the $\alpha$ parameter, and therefore the ones that are  included in \eqref{WGCqforms}, are also limited (see e.g. \cite{Gendler:2020dfp} for a detailed discussion on this issue for the context of particles). In fact, they are restricted to the cases where the brane-scalar interaction, which is given by the derivative of the tension of the brane with respect to the field, is proportional to the tension itself.

\section{The 4d $\mathcal{N}=1$ Scalar Potential in terms of Membranes}
\label{section3}

The main goal of this note is to point out a correspondence between the different terms of the 4d $\mathcal{N}=1$ F-term scalar potential and the interaction between two BPS membranes due to particle exchange.This scalar potential takes the following form in terms of the K\"ahler potential, $K$ and the superpotential, $W$ \cite{sugra}
\begin{equation}
\label{Cremmeretal}
    V=e^K\left( K^{I \bar{J}} D_I W D_{\bar{J}}\bar{W}- 3 |W|^2\right),
\end{equation}
where the K\"ahler covariant derivative is given by $D_I W= \partial_I W + W \partial _I K$ and the $I, J$ indices run over all the complex scalar fields of the theory. 

\subsection{The Scalar Potential and the 3-form Interaction}
In order to make the correspondence precise, the first important ingredient is the fact that  in type II flux compactifications, this scalar potential can also be expressed as the following bilinear \cite{Bielleman:2015ina, Herraez:2018vae} 
\begin{equation}
V=\dfrac{1}{2} Z^{AB} Q_A Q_B,
\end{equation}
where $Q_A$ is a vector containing the fluxes, and $Z^{AB}$ is the inverse of the matrix that appears in the kinetic term of 3-forms, which is of the form $\mathcal{S}_{\mathrm{3, \, kin}}= \frac{1}{2 \kappa_4^2} \int Z_{AB} F^A \wedge \star F^B $. This matrix encodes the dependence on the scalar fields (both the axions and the saxions), and the charge vectors encode all the information about the fluxes.  This form of the potential has been argued to be valid fore more general $\mathcal{N}=1$ setups in \cite{Farakos:2017jme,Farakos:2017ocw,Bandos:2018gjp,Bandos:2019wgy,Lanza:2019xxg} and this is expected to be quite general at least at the level of any 4d EFT, since a cosmological constant term can be always rewritten in terms of 4-forms, so allowing for field dependent kinetic terms seems to be enough to encode this kind of scalar potentials.\footnote{There is an important caveat for this, namely the fact that the matrix $Z_{AB}$ needs to be invertible in order for one to be able to write the potential in terms of 4-forms.} This bilinear expression is suggestive, since it resembles the form of an electric interaction between two charged objects, with charges $Q_A$, mediated by some mediator with propagator $\sim Z^{AB}$. In fact, this is the case when we consider the 3-form interaction between two flat membranes, whose couplings with the different 3-forms on the spectrum are given by a generalization of eq. \eqref{SCS}. The propagator of the 3-forms takes the form
\begin{equation}
 \langle A^{A, \, \mu \nu \rho}\,  A^B_{\alpha \beta \gamma}  \rangle =i \, 3! \, \dfrac{ \kappa_4^2  }{k^2}  \,Z^{AB} \,  \delta_{[\alpha}^{\mu} \delta_{\beta}^{\nu} \delta_{\gamma] }^{\rho}  \, ,
\end{equation}
and the vertex between the membrane and the 3-form is the straightforward generalization of  eq. \eqref{fig:qformvertex}, namely
\begin{equation}
\includegraphics[align=c, height=54pt]{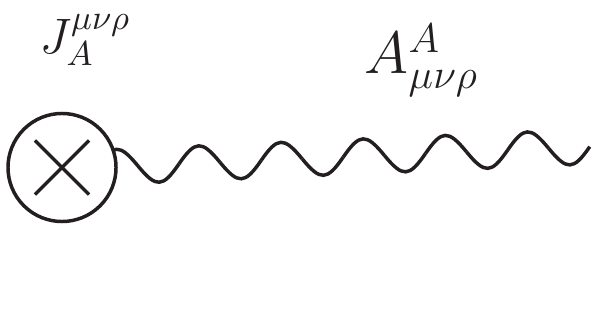}
      \  = \  i \dfrac{Q_A}{3!}\epsilon^{a b c} \delta_{a}^{[\mu}\delta_{b}^{\nu}\delta_{c}^{\rho]} \, .
        \label{fig:3formvertex} 
\end{equation}

\begin{figure}[tb]
  \label{fig:3formexchange}
   \begin{center}
 \includegraphics[align=c, height=48pt]{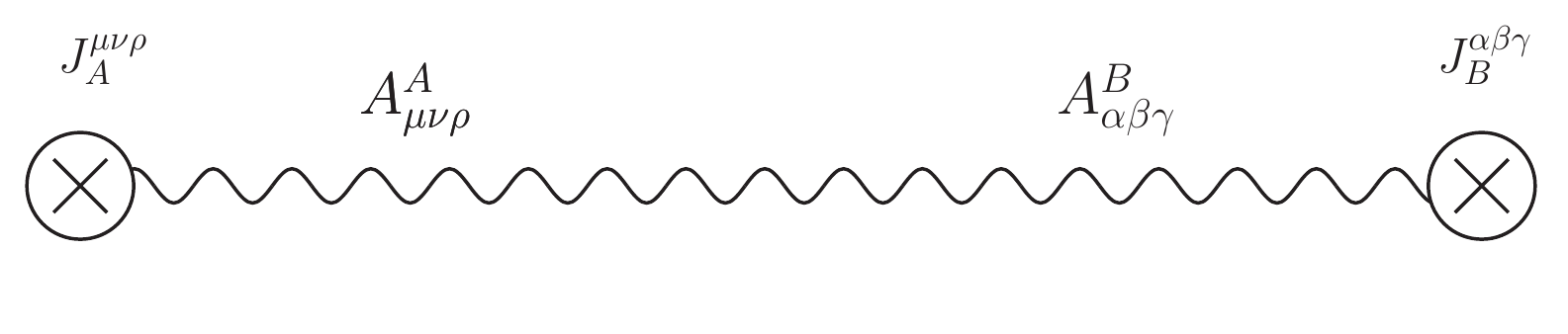}
      
          \caption{Diagramatic representation of the tree-level $3$-form exchange between two membranes}  
     \end{center}
\end{figure}
Then the diagram in fig. \ref{fig:3formexchange} gives an amplitude
\begin{equation}
\mathcal{A}_{3-\mathrm{form}}=-\dfrac{ \kappa_4^2 |\bar{g}_{ab}|  }{k^2} \,  Z^{AB} Q_A Q_B,
\end{equation}
which is proportional to the potential, as announced. 

\subsection{The Scalar and Graviton Interaction}
We turn now to the computation of the diagrams corresponding to the exchange of gravitons and scalars. For flat BPS membranes, which preserve 1/2 of the supersymmetries of the Minkowski background along their worldvolume, the tensions takes the form \cite{Bandos:2018gjp,Bandos:2019wgy,Font:2019cxq, Lanza:2019xxg}
\begin{equation}
\label{TensionBPS}
T=2e^{K/2} |W|
\end{equation}
where, $W$ corresponds to the superpotential sourced by the membrane \footnote{Notice that in our case we are considering a membrane separating a region where the superpotential vanishes identically from another one where the superpotential is generated by the membrane, so that the change in the superpotential is the superpotential itself.} and can be directly related to its  charges $Q_A$ and the so-called period vector. In type II we can interpret these membranes as coming from bound states of D$p$ or NS5-branes wrapping internal cycles and their charges are related to the fluxes at the other side of the membrane.
 
The graviton exchange between two flat equal membranes of tension $T_p$  is universal and for codimension 1 objects is given by the second term in eq. \eqref{As+g}. Upon substitution of eq. \eqref{TensionBPS} and $D=4$  we find
\begin{equation}
\mathcal{A}_g=- \dfrac{2 \kappa_4^2 |g_{ab}|}{k^2}\, 3e^K |W|^2 \, . 
\end{equation}

In order to calculate the contribution from the complex scalars, we recall that in a 4d $\mathcal{N}=1$ theory they have the following kinetic terms
\begin{equation}
    S_{\Phi,\, \mathrm{kin }}=\dfrac{1}{\kappa^2_4}\int \sqrt{-g}\, d^4 x \, K_{I\bar{J}}\partial_\mu \Phi^I \partial^\mu \bar{\Phi}^{\bar{J}},
\end{equation}
and we call $\preal{(\Phi^I)}=X^I$ to the saxionic part and $\pim{ (\Phi^I)}=Y^I$ to the axionic one, with propagators (in momentum space)
\begin{equation}
    \langle X^I X^J  \rangle=  -\dfrac{i\kappa_4^2}{2 k^2}  K^{I\bar{J}}, \qquad 
     \langle Y^I Y^J  \rangle=  -\dfrac{i\kappa_4^2}{2 k^2}  K^{I\bar{J}}.
\end{equation}

Additionally, the coupling between the scalars and the membrane can be obtained from the action \cite{Bandos:2018gjp,Bandos:2019wgy,Font:2019cxq, Lanza:2019xxg}
\begin{equation}
    S_{mem}= -\int_{WV} \, d^3\xi \, \sqrt{|g_{ab}|} \, T\left( \Phi^I, \bar{\Phi}^{\bar{J}} \right) + S_{CS},
\end{equation}
where the tension is given by eq. \eqref{TensionBPS}. The scalar-membrane vertex is obtained from the first derivative of the tension with respect to each of the (real) fields evaluated in the background. In order to relate this with eq. \eqref{Cremmeretal}, we need to express everything in terms of the complex fields. To do so, we use the fact that $K=K(\Phi^I + \bar{\Phi}^{\bar{J}})$ is a real function that depends only on the saxions $X^I$, and that $W=W(\Phi)$ is a holomorphic function in order to re-express the derivatives of the tension with respect to $X^I$ and $Y^I$ in terms of derivatives with respect to $Z^I$ and $\bar{Z}^I$. From now on we restrict to a single complex scalar field $Z=X+iY$ for simplicity, but the generalization to more fields is straightforward. 

\begin{figure}[tb]

    \begin{center}
     	  \includegraphics[align=c, height=40pt]{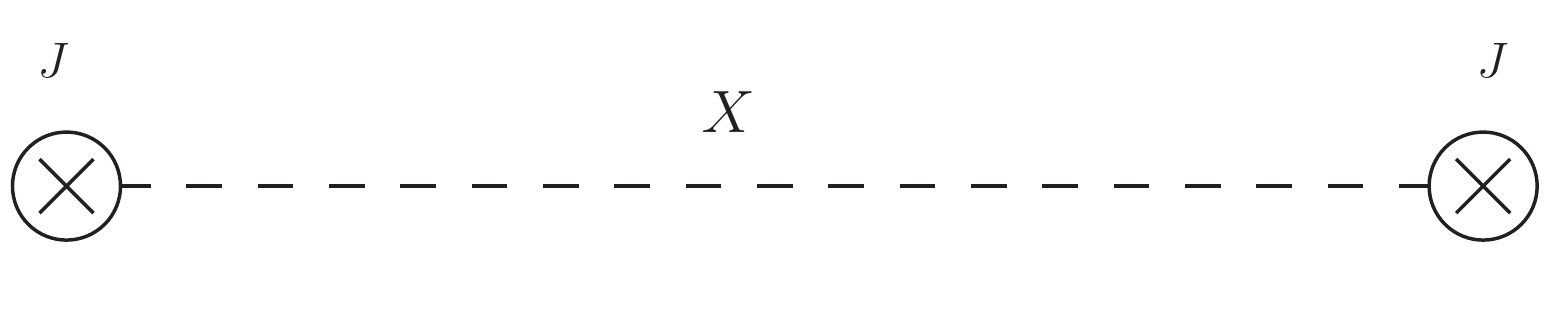}
        \label{fig:scalarexchangeX}
$+$
        \includegraphics[align=c, height=40pt]{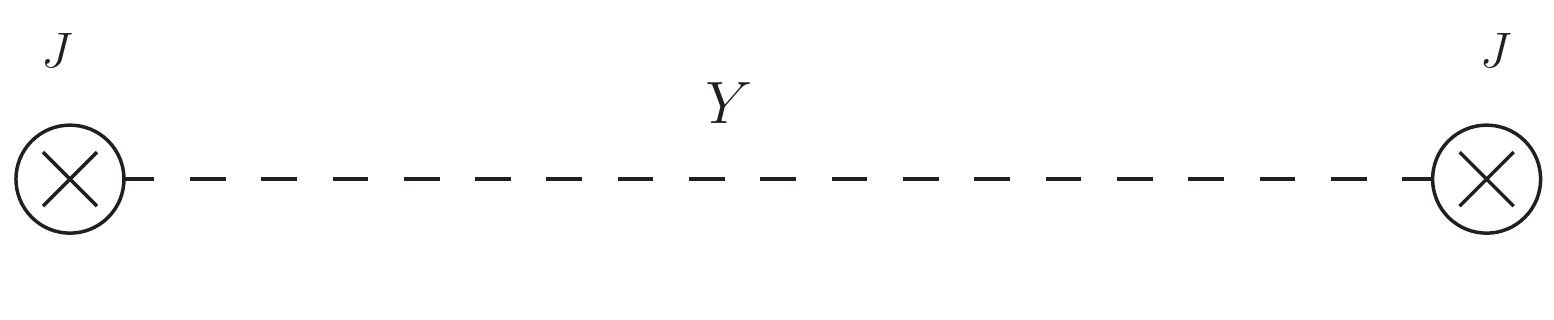}
        \label{fig:scalarexchangeY}
        \caption{Diagramatic representation of the tree-level scalar exchange corresponding to the real and imaginary parts of a complex scalar}
       \label{fig:ReImscalars}
     \end{center}
\end{figure}  

  The vertices we are interested in take the form
  
  \begin{equation}
      \begin{split}
        \includegraphics[align=c, height=48pt]{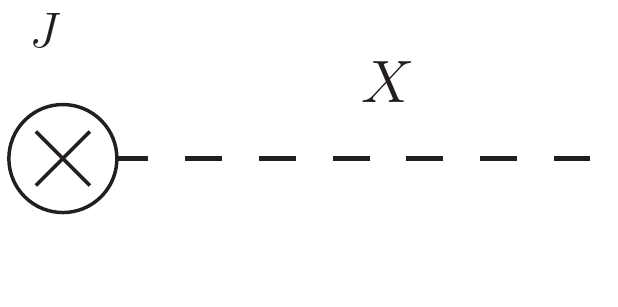}
        \label{fig:membraneX} =\,  2\sqrt{|g_{ab}|}\dfrac{ \partial T}{\partial X}\, = & \, i \sqrt{|g_{ab}|}\,  e^{K/2}\left[ \dfrac{1}{2} \dfrac{\partial K}{\partial X} + \dfrac{1}{2 |W|} \left( \dfrac{ \partial W}{\partial X} \bar{W}+ \dfrac{ \partial \bar{W}}{\partial X} W \right) \right]= \\
         = &\, 2i \sqrt{|g_{ab}|} \, e^{K/2}\left[ (\partial_Z K) |W|+ \dfrac{1}{ |W|}  \preal{\left[(\partial_Z W) \bar{W} \right]} \right]
      \end{split}
  \end{equation}
  
\begin{equation}
      \begin{split}
      \includegraphics[align=c, height=48pt]{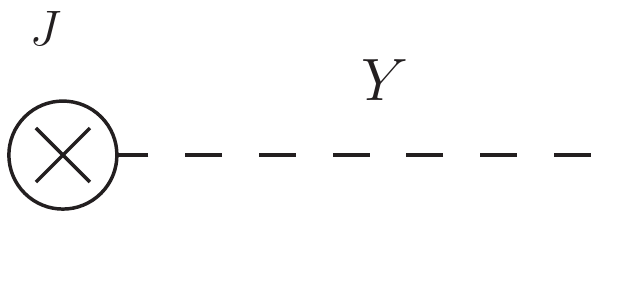}
        \label{fig:membraneY} =\,  2i\sqrt{|g_{ab}|} \dfrac{ \partial T}{\partial Y}\, = & \, i\sqrt{|g_{ab}|}\,  e^{K/2}\left[  \dfrac{1}{2 |W|} \left( \dfrac{ \partial W}{\partial Y} \bar{W}+ \dfrac{ \partial \bar{W}}{\partial Y} W \right) \right]= \\
         = &\,  -2 i \sqrt{|g_{ab}|}\,  e^{K/2} \dfrac{1}{ |W|}  \pim {\left[(\partial_Z W) \bar{W} \right]}.
      \end{split}
  \end{equation}
    
Having calculated the vertices and the propagators we can calculate the amplitude associated to scalar exchange, in fig. \ref{fig:ReImscalars}, which yields
\begin{equation}
    \mathcal{A}_{X+Y}=  \dfrac{2 \kappa_4^2 |g_{ab}|}{ k^2} K^{Z\bar{Z}}\left[ \left( \dfrac{\partial T}{\partial X}\right)^2+ \left( \dfrac{\partial T}{\partial Y}\right)^2 \right] \, = \, \dfrac{2 \kappa_4^2 |g_{ab}|}{k^2} e^K  (K^{Z\bar{Z}} D_Z W D_{\bar{Z}}\bar{W})\, .
\end{equation}
The interaction between the membranes then cancels if
\begin{equation}
\label{Cremmeretal2}
\dfrac{1}{2}Z^{AB}Q_AQ_B\, =\, e^KK^{I \bar{J}} D_I W D_{\bar{J}}\bar{W}- 3 e^K |W|^2 \, .
\end{equation}
There is therefore a one-to-one correspondence between each of the three  interactions of the membranes and eq. \eqref{Cremmeretal}.  In the membrane picture,the 3-form interactions correspond to the potential in \eqref{Cremmeretal} and they equal the scalar and graviton interactions, which correspond to the two terms on the r.h.s. of eq. \eqref{Cremmeretal}, respectively. The upshot is that for every point in the scalar field space, we have a picture in which there is a potential, and another one with two BPS membranes in a Minkowski background which feel no net self-interactions. Therefore, whereas in the first picture the potential gets a contribution from the K\"ahler covariant derivatives of the superpotential and another one from the $-3|W|^2$ term,  in the membrane picture these match the scalar and graviton interactions, respectively. In this language, for example, a supersymmetric vacuum of the potential would correspond to a pair of membranes with vanishing scalar interaction. Notice, moreover, that in terms of the scalar potential description, this correspondence is valid off-shell, since it is defined for every point in scalar field space, not only for the vacua of the potential. 

Besides, in the membrane picture, the background is always Minkowski. This is the case because between the two membranes, the cosmological constant contribution (sourced by the membranes themselves) is  encoded in the 3-form interaction, whose energy density between the membranes is canceled by the scalar and graviton interaction, resulting in a vanishing energy density between the membranes due to the interaction with each other. 

\section{Summary and Outlook}
\label{s:summary}
To sum up, we have studied the tree-level interaction between $p$-branes due to exchange of scalars, gravitons and $(p+1)$-forms.  We have shown that, for the particular case of BPS membranes in a 4d Minkowski background, there is a correspondence between each interaction and each term in the $\mathcal{N}=1$ scalar potential. The fact that the scalar potential can be written in that form may be translated in the membrane picture to the requirement that the net force between the two membranes vanishes, that is, that the 3-form interaction cancels the scalar plus graviton contributions. From the point of view of the potential, this correspondence is valid off-shell (not only in the minima). This means that for every point in the scalar field space, characterized by a value of the scalar potential, there exists a corresponding membrane configuration with  the same values for the scalar fields in which the self-interaction vanishes and whose 3-form interaction, or equivalently the scalar plus graviton interaction, equals the value of the potential in the initial picture. Let us remark that we have only worked out in detail the 4d case, but we expect similar arguments to apply for a relation between codimension 1 BPS objects  and scalar potentials in higher dimensions.

This correspondence, although interesting per se, might be useful for the swampland program (see  \cite{PaltiTalkSP19, LMMV}). In that context, lots of recent results suggest a very intricate web of swampland conjectures, in which apparently disconnected conjectures happen to be related or even imply each other in many different ways (see e.g. \cite{review, timo2, Gendler:2020dfp, Andriot:2020lea} and references therein). In this respect, some swampland conjectures, like the WGC  or the SDC  make statements about the properties of the spectrum of consistent theories of QG, whereas others like the dS Conjecture  or the ADC refer to the kinds of potentials or vacua that are allowed in QG. We believe that the correspondence explained in this note could help to uncover some connections between these two apparently different types of statements, since it relates configurations with a scalar potential with configurations of charged extended objects.

\bigskip

\centerline{\bf \large Acknowledgments}

\bigskip

\noindent The author would like to thank L. Ib\'a\~nez, E. Palti and I. Valenzuela for useful discussions. 
This work is supported by the Spanish Research Agency (Agencia Estatal de Investigacion) through the grant IFT 
Centro de Excelencia Severo Ochoa SEV-2016-0597, and by the grants 
FPA2015-65480-P and PGC2018-095976-B-C21 from MCIU/AEI/FEDER, UE. 
The author is supported by the Spanish FPU Grant No. FPU15/05012.

\newpage
\bibliographystyle{JHEP}
\inputencoding{latin2}
\bibliography{alvarobib}

\providecommand{\href}[2]{#2}\begingroup\raggedright\begin{thebibliography}{10}

\bibitem{pioneros1}
M.~Duff and P.~van Nieuwenhuizen, \emph{{Quantum Inequivalence of Different
  Field Representations}},
  \href{https://doi.org/10.1016/0370-2693(80)90852-7}{\emph{Phys. Lett. B}
  {\bfseries 94} (1980) 179}.

\bibitem{pioneros2}
S.~Hawking, \emph{{The Cosmological Constant Is Probably Zero}},
  \href{https://doi.org/10.1016/0370-2693(84)91370-4}{\emph{Phys. Lett. B}
  {\bfseries 134} (1984) 403}.

\bibitem{pioneros3}
M.~J. Duff, \emph{{The Cosmological Constant Is Possibly Zero, but the Proof Is
  Probably Wrong}},
  \href{https://doi.org/10.1016/0370-2693(89)90284-0}{\emph{Phys. Lett. B}
  {\bfseries 226} (1989) 36}.

\bibitem{pioneros4}
Z.~C. Wu, \emph{{The Cosmological Constant is Probably Zero, and a Proof is
  Possibly Right}},
  \href{https://doi.org/10.1016/j.physletb.2007.12.019}{\emph{Phys. Lett. B}
  {\bfseries 659} (2008) 891}
  [\href{https://arxiv.org/abs/0709.3314}{{\ttfamily 0709.3314}}].

\bibitem{pioneros5}
M.~J. Duncan and L.~G. Jensen, \emph{{Four Forms and the Vanishing of the
  Cosmological Constant}},
  \href{https://doi.org/10.1016/0550-3213(90)90344-D}{\emph{Nucl. Phys. B}
  {\bfseries 336} (1990) 100}.

\bibitem{BT}
J.~Brown and C.~Teitelboim, \emph{{Dynamical Neutralization of the Cosmological
  Constant}}, \href{https://doi.org/10.1016/0370-2693(87)91190-7}{\emph{Phys.
  Lett. B} {\bfseries 195} (1987) 177}.

\bibitem{BT2}
J.~Brown and C.~Teitelboim, \emph{{Neutralization of the Cosmological Constant
  by Membrane Creation}},
  \href{https://doi.org/10.1016/0550-3213(88)90559-7}{\emph{Nucl. Phys. B}
  {\bfseries 297} (1988) 787}.

\bibitem{BP}
R.~Bousso and J.~Polchinski, \emph{{Quantization of four form fluxes and
  dynamical neutralization of the cosmological constant}},
  \href{https://doi.org/10.1088/1126-6708/2000/06/006}{\emph{JHEP} {\bfseries
  06} (2000) 006} [\href{https://arxiv.org/abs/hep-th/0004134}{{\ttfamily
  hep-th/0004134}}].

\bibitem{morerecent2}
G.~Dvali, \emph{{Large hierarchies from attractor vacua}},
  \href{https://doi.org/10.1103/PhysRevD.74.025018}{\emph{Phys. Rev. D}
  {\bfseries 74} (2006) 025018}
  [\href{https://arxiv.org/abs/hep-th/0410286}{{\ttfamily hep-th/0410286}}].

\bibitem{morerecent3}
G.~Dvali, \emph{{Three-form gauging of axion symmetries and gravity}},
  \href{https://arxiv.org/abs/hep-th/0507215}{{\ttfamily hep-th/0507215}}.

\bibitem{morerecent4}
G.~Dvali, \emph{{A Vacuum accumulation solution to the strong CP problem}},
  \href{https://doi.org/10.1103/PhysRevD.74.025019}{\emph{Phys. Rev. D}
  {\bfseries 74} (2006) 025019}
  [\href{https://arxiv.org/abs/hep-th/0510053}{{\ttfamily hep-th/0510053}}].

\bibitem{morerecent5}
G.~Dvali, S.~Folkerts and A.~Franca, \emph{{How neutrino protects the axion}},
  \href{https://doi.org/10.1103/PhysRevD.89.105025}{\emph{Phys. Rev. D}
  {\bfseries 89} (2014) 105025}
  [\href{https://arxiv.org/abs/1312.7273}{{\ttfamily 1312.7273}}].

\bibitem{KS}
N.~Kaloper and L.~Sorbo, \emph{{A Natural Framework for Chaotic Inflation}},
  \href{https://doi.org/10.1103/PhysRevLett.102.121301}{\emph{Phys. Rev. Lett.}
  {\bfseries 102} (2009) 121301}
  [\href{https://arxiv.org/abs/0811.1989}{{\ttfamily 0811.1989}}].

\bibitem{KLS}
N.~Kaloper, A.~Lawrence and L.~Sorbo, \emph{{An Ignoble Approach to Large Field
  Inflation}}, \href{https://doi.org/10.1088/1475-7516/2011/03/023}{\emph{JCAP}
  {\bfseries 03} (2011) 023} [\href{https://arxiv.org/abs/1101.0026}{{\ttfamily
  1101.0026}}].

\bibitem{Dudas:2014pva}
E.~Dudas, \emph{{Three-form multiplet and Inflation}},
  \href{https://doi.org/10.1007/JHEP12(2014)014}{\emph{JHEP} {\bfseries 12}
  (2014) 014} [\href{https://arxiv.org/abs/1407.5688}{{\ttfamily 1407.5688}}].

\bibitem{Escobar:2015ckf}
D.~Escobar, A.~Landete, F.~Marchesano and D.~Regalado, \emph{{D6-branes and
  axion monodromy inflation}},
  \href{https://doi.org/10.1007/JHEP03(2016)113}{\emph{JHEP} {\bfseries 03}
  (2016) 113} [\href{https://arxiv.org/abs/1511.08820}{{\ttfamily
  1511.08820}}].

\bibitem{imuv}
L.~E. Ib\'a\~nez, M.~Montero, A.~Uranga and I.~Valenzuela, \emph{{Relaxion
  Monodromy and the Weak Gravity Conjecture}},
  \href{https://doi.org/10.1007/JHEP04(2016)020}{\emph{JHEP} {\bfseries 04}
  (2016) 020} [\href{https://arxiv.org/abs/1512.00025}{{\ttfamily
  1512.00025}}].

\bibitem{Carta:2016ynn}
F.~Carta, F.~Marchesano, W.~Staessens and G.~Zoccarato, \emph{{Open string
  multi-branched and K\"ahler potentials}},
  \href{https://doi.org/10.1007/JHEP09(2016)062}{\emph{JHEP} {\bfseries 09}
  (2016) 062} [\href{https://arxiv.org/abs/1606.00508}{{\ttfamily
  1606.00508}}].

\bibitem{Garcia_Valdecasas:2016voz}
E.~Garc\'ia-Valdecasas and A.~Uranga, \emph{{On the 3-form formulation of axion
  potentials from D-brane instantons}},
  \href{https://doi.org/10.1007/JHEP02(2017)087}{\emph{JHEP} {\bfseries 02}
  (2017) 087} [\href{https://arxiv.org/abs/1605.08092}{{\ttfamily
  1605.08092}}].

\bibitem{Valenzuela:2016yny}
I.~Valenzuela, \emph{{Backreaction Issues in Axion Monodromy and Minkowski
  4-forms}}, \href{https://doi.org/10.1007/JHEP06(2017)098}{\emph{JHEP}
  {\bfseries 06} (2017) 098}
  [\href{https://arxiv.org/abs/1611.00394}{{\ttfamily 1611.00394}}].

\bibitem{Blumenhagen:2017cxt}
R.~Blumenhagen, I.~Valenzuela and F.~Wolf, \emph{{The Swampland Conjecture and
  F-term Axion Monodromy Inflation}},
  \href{https://doi.org/10.1007/JHEP07(2017)145}{\emph{JHEP} {\bfseries 07}
  (2017) 145} [\href{https://arxiv.org/abs/1703.05776}{{\ttfamily
  1703.05776}}].

\bibitem{Bielleman:2015ina}
S.~Bielleman, L.~E. Ib\'a\~nez and I.~Valenzuela, \emph{{Minkowski 3-forms,
  Flux String Vacua, Axion Stability and Naturalness}},
  \href{https://doi.org/10.1007/JHEP12(2015)119}{\emph{JHEP} {\bfseries 12}
  (2015) 119} [\href{https://arxiv.org/abs/1507.06793}{{\ttfamily
  1507.06793}}].

\bibitem{Herraez:2018vae}
A.~Herraez, L.~E. Ib\'a\~nez, F.~Marchesano and G.~Zoccarato, \emph{{The Type
  IIA Flux Potential, 4-forms and Freed-Witten anomalies}},
  \href{https://doi.org/10.1007/JHEP09(2018)018}{\emph{JHEP} {\bfseries 09}
  (2018) 018} [\href{https://arxiv.org/abs/1802.05771}{{\ttfamily
  1802.05771}}].

\bibitem{swampland}
C.~Vafa, \emph{{The String landscape and the swampland}},
  \href{https://arxiv.org/abs/hep-th/0509212}{{\ttfamily hep-th/0509212}}.

\bibitem{vafafederico}
T.~D. Brennan, F.~Carta and C.~Vafa, \emph{{The String Landscape, the
  Swampland, and the Missing Corner}},
  \href{https://doi.org/10.22323/1.305.0015}{\emph{PoS} {\bfseries TASI2017}
  (2017) 015} [\href{https://arxiv.org/abs/1711.00864}{{\ttfamily
  1711.00864}}].

\bibitem{review}
E.~Palti, \emph{{The Swampland: Introduction and Review}},
  \href{https://doi.org/10.1002/prop.201900037}{\emph{Fortsch. Phys.}
  {\bfseries 67} (2019) 1900037}
  [\href{https://arxiv.org/abs/1903.06239}{{\ttfamily 1903.06239}}].

\bibitem{dS1}
G.~Obied, H.~Ooguri, L.~Spodyneiko and C.~Vafa, \emph{{De Sitter Space and the
  Swampland}},  \href{https://arxiv.org/abs/1806.08362}{{\ttfamily
  1806.08362}}.

\bibitem{Krishnan}
S.~K. Garg and C.~Krishnan, \emph{{Bounds on Slow Roll and the de Sitter
  Swampland}}, \href{https://doi.org/10.1007/JHEP11(2019)075}{\emph{JHEP}
  {\bfseries 11} (2019) 075}
  [\href{https://arxiv.org/abs/1807.05193}{{\ttfamily 1807.05193}}].

\bibitem{dS3}
H.~Ooguri, E.~Palti, G.~Shiu and C.~Vafa, \emph{{Distance and de Sitter
  Conjectures on the Swampland}},
  \href{https://doi.org/10.1016/j.physletb.2018.11.018}{\emph{Phys. Lett. B}
  {\bfseries 788} (2019) 180}
  [\href{https://arxiv.org/abs/1810.05506}{{\ttfamily 1810.05506}}].

\bibitem{TCC}
A.~Bedroya and C.~Vafa, \emph{{Trans-Planckian Censorship and the Swampland}},
  \href{https://arxiv.org/abs/1909.11063}{{\ttfamily 1909.11063}}.

\bibitem{Ooguri:2016pdq}
H.~Ooguri and C.~Vafa, \emph{{Non-supersymmetric AdS and the Swampland}},
  \href{https://doi.org/10.4310/ATMP.2017.v21.n7.a8}{\emph{Adv. Theor. Math.
  Phys.} {\bfseries 21} (2017) 1787}
  [\href{https://arxiv.org/abs/1610.01533}{{\ttfamily 1610.01533}}].

\bibitem{lpv}
D.~L\"ust, E.~Palti and C.~Vafa, \emph{{AdS and the Swampland}},
  \href{https://doi.org/10.1016/j.physletb.2019.134867}{\emph{Phys. Lett. B}
  {\bfseries 797} (2019) 134867}
  [\href{https://arxiv.org/abs/1906.05225}{{\ttfamily 1906.05225}}].

\bibitem{Gautason:2015tig}
F.~Gautason, M.~Schillo, T.~Van~Riet and M.~Williams, \emph{{Remarks on scale
  separation in flux vacua}},
  \href{https://doi.org/10.1007/JHEP03(2016)061}{\emph{JHEP} {\bfseries 03}
  (2016) 061} [\href{https://arxiv.org/abs/1512.00457}{{\ttfamily
  1512.00457}}].

\bibitem{WGC}
N.~Arkani-Hamed, L.~Motl, A.~Nicolis and C.~Vafa, \emph{{The String landscape,
  black holes and gravity as the weakest force}},
  \href{https://doi.org/10.1088/1126-6708/2007/06/060}{\emph{JHEP} {\bfseries
  06} (2007) 060} [\href{https://arxiv.org/abs/hep-th/0601001}{{\ttfamily
  hep-th/0601001}}].

\bibitem{distance}
H.~Ooguri and C.~Vafa, \emph{{On the Geometry of the String Landscape and the
  Swampland}},
  \href{https://doi.org/10.1016/j.nuclphysb.2006.10.033}{\emph{Nucl. Phys. B}
  {\bfseries 766} (2007) 21}
  [\href{https://arxiv.org/abs/hep-th/0605264}{{\ttfamily hep-th/0605264}}].

\bibitem{timo2}
S.-J. Lee, W.~Lerche and T.~Weigand, \emph{{A Stringy Test of the Scalar Weak
  Gravity Conjecture}},
  \href{https://doi.org/10.1016/j.nuclphysb.2018.11.001}{\emph{Nucl. Phys. B}
  {\bfseries 938} (2019) 321}
  [\href{https://arxiv.org/abs/1810.05169}{{\ttfamily 1810.05169}}].

\bibitem{Gendler:2020dfp}
N.~Gendler and I.~Valenzuela, \emph{{Merging the Weak Gravity and Distance
  Conjectures Using BPS Extremal Black Holes}},
  \href{https://arxiv.org/abs/2004.10768}{{\ttfamily 2004.10768}}.

\bibitem{Andriot:2020lea}
D.~Andriot, N.~Cribiori and D.~Erkinger, \emph{{The web of swampland
  conjectures and the TCC bound}},
  \href{https://arxiv.org/abs/2004.00030}{{\ttfamily 2004.00030}}.

\bibitem{PaltiTalkSP19}
E.~Palti, \emph{{Distance Conjecture and Potentials}}. Talk at String
  Phenomenology 2019, CERN.

\bibitem{LMMV}
S.~Lanza, F.~Marchesano, L.~Martucci and I.~Valenzuela, \emph{{Swampland
  Conjectures for Strings and Membranes}},
  \href{https://arxiv.org/abs/2006.15154}{{\ttfamily 2006.15154}}.

\bibitem{Polchinski:1998rq}
J.~Polchinski, \emph{{String theory. Vol. 1: An introduction to the bosonic
  string}}, Cambridge Monographs on Mathematical Physics. Cambridge University
  Press, 12, 2007,
  \href{https://doi.org/10.1017/CBO9780511816079}{10.1017/CBO9780511816079}.

\bibitem{Polchinski:1998rr}
J.~Polchinski, \emph{{String theory. Vol. 2: Superstring theory and beyond}},
  Cambridge Monographs on Mathematical Physics. Cambridge University Press, 12,
  2007,
  \href{https://doi.org/10.1017/CBO9780511618123}{10.1017/CBO9780511618123}.

\bibitem{Polchinski:1995mt}
J.~Polchinski, \emph{{Dirichlet Branes and Ramond-Ramond charges}},
  \href{https://doi.org/10.1103/PhysRevLett.75.4724}{\emph{Phys. Rev. Lett.}
  {\bfseries 75} (1995) 4724}
  [\href{https://arxiv.org/abs/hep-th/9510017}{{\ttfamily hep-th/9510017}}].

\bibitem{BOOK}
L.~E. Ib{\'{a}}{\~{n}}ez and A.~M. Uranga, \emph{String Theory and Particle
  Physics: An Introduction to String Phenomenology}. Cambridge University
  Press, 2012,
  \href{https://doi.org/10.1017/CBO9781139018951}{10.1017/CBO9781139018951}.

\bibitem{Alvarez:2020zul}
E.~Alvarez, \emph{{Windows on Quantum Gravity}},  5, 2020,
  \href{https://arxiv.org/abs/2005.09466}{{\ttfamily 2005.09466}}.

\bibitem{Luscher}
M.~Luscher, \emph{{The Secret Long Range Force in Quantum Field Theories With
  Instantons}}, \href{https://doi.org/10.1016/0370-2693(78)90487-2}{\emph{Phys.
  Lett. B} {\bfseries 78} (1978) 465}.

\bibitem{WGC16}
B.~Heidenreich, M.~Reece and T.~Rudelius, \emph{{Sharpening the Weak Gravity
  Conjecture with Dimensional Reduction}},
  \href{https://doi.org/10.1007/JHEP02(2016)140}{\emph{JHEP} {\bfseries 02}
  (2016) 140} [\href{https://arxiv.org/abs/1509.06374}{{\ttfamily
  1509.06374}}].

\bibitem{Palti}
E.~Palti, \emph{{The Weak Gravity Conjecture and Scalar Fields}},
  \href{https://doi.org/10.1007/JHEP08(2017)034}{\emph{JHEP} {\bfseries 08}
  (2017) 034} [\href{https://arxiv.org/abs/1705.04328}{{\ttfamily
  1705.04328}}].

\bibitem{Heidenreich:2019zkl}
B.~Heidenreich, M.~Reece and T.~Rudelius, \emph{{Repulsive Forces and the Weak
  Gravity Conjecture}},
  \href{https://doi.org/10.1007/JHEP10(2019)055}{\emph{JHEP} {\bfseries 10}
  (2019) 055} [\href{https://arxiv.org/abs/1906.02206}{{\ttfamily
  1906.02206}}].

\bibitem{Horowitz:1991cd}
G.~T. Horowitz and A.~Strominger, \emph{{Black strings and P-branes}},
  \href{https://doi.org/10.1016/0550-3213(91)90440-9}{\emph{Nucl. Phys. B}
  {\bfseries 360} (1991) 197}.

\bibitem{Gonzalo:2020kke}
E.~Gonzalo and L.~E. Ib\'a\~nez, \emph{{Pair Production and Gravity as the
  Weakest Force}},  \href{https://arxiv.org/abs/2005.07720}{{\ttfamily
  2005.07720}}.

\bibitem{irene}
T.~W. Grimm, E.~Palti and I.~Valenzuela, \emph{{Infinite Distances in Field
  Space and Massless Towers of States}},
  \href{https://doi.org/10.1007/JHEP08(2018)143}{\emph{JHEP} {\bfseries 08}
  (2018) 143} [\href{https://arxiv.org/abs/1802.08264}{{\ttfamily
  1802.08264}}].

\bibitem{Font:2019cxq}
A.~Font, A.~Herraez and L.~E. Ib\'a\~nez, \emph{{The Swampland Distance
  Conjecture and Towers of Tensionless Branes}},
  \href{https://doi.org/10.1007/JHEP08(2019)044}{\emph{JHEP} {\bfseries 08}
  (2019) 044} [\href{https://arxiv.org/abs/1904.05379}{{\ttfamily
  1904.05379}}].

\bibitem{sugra}
D.~Z. Freedman and A.~Van~Proeyen, \emph{{Supergravity}}. Cambridge Univ.
  Press, Cambridge, UK, 5, 2012.

\bibitem{Farakos:2017jme}
F.~Farakos, S.~Lanza, L.~Martucci and D.~Sorokin, \emph{{Three-forms in
  Supergravity and Flux Compactifications}},
  \href{https://doi.org/10.1140/epjc/s10052-017-5185-y}{\emph{Eur. Phys. J. C}
  {\bfseries 77} (2017) 602}
  [\href{https://arxiv.org/abs/1706.09422}{{\ttfamily 1706.09422}}].

\bibitem{Farakos:2017ocw}
F.~Farakos, S.~Lanza, L.~Martucci and D.~Sorokin, \emph{{Three-forms,
  Supersymmetry and String Compactifications}},
  \href{https://doi.org/10.1134/S1063779618050192}{\emph{Phys. Part. Nucl.}
  {\bfseries 49} (2018) 823}
  [\href{https://arxiv.org/abs/1712.09366}{{\ttfamily 1712.09366}}].

\bibitem{Bandos:2018gjp}
I.~Bandos, F.~Farakos, S.~Lanza, L.~Martucci and D.~Sorokin,
  \emph{{Three-forms, dualities and membranes in four-dimensional
  supergravity}}, \href{https://doi.org/10.1007/JHEP07(2018)028}{\emph{JHEP}
  {\bfseries 07} (2018) 028}
  [\href{https://arxiv.org/abs/1803.01405}{{\ttfamily 1803.01405}}].

\bibitem{Bandos:2019wgy}
I.~Bandos, F.~Farakos, S.~Lanza, L.~Martucci and D.~Sorokin, \emph{{Higher
  Forms and Membranes in 4D Supergravities}},
  \href{https://doi.org/10.1002/prop.201910020}{\emph{Fortsch. Phys.}
  {\bfseries 67} (2019) 1910020}
  [\href{https://arxiv.org/abs/1903.02841}{{\ttfamily 1903.02841}}].

\bibitem{Lanza:2019xxg}
S.~Lanza, F.~Marchesano, L.~Martucci and D.~Sorokin, \emph{{How many fluxes fit
  in an EFT?}}, \href{https://doi.org/10.1007/JHEP10(2019)110}{\emph{JHEP}
  {\bfseries 10} (2019) 110}
  [\href{https://arxiv.org/abs/1907.11256}{{\ttfamily 1907.11256}}].

\end{thebibliography}\endgroup
\inputencoding{utf8}

\end{document}